\documentclass{article}[12pt]
\usepackage{float}
\usepackage{amsfonts}
\usepackage{graphicx}
\usepackage{amsmath}
\usepackage{appendix}
\usepackage{fullpage}
\usepackage{multicol}
\usepackage{hyperref}
\hypersetup{
  colorlinks   = true, 
  urlcolor     = blue, 
  linkcolor    = blue, 
  citecolor    = blue  
}

\newtheorem{remark}{Remark}
\newtheorem{theorem}{Theorem}
\newenvironment{proof}[1][Proof]{\begin{trivlist}
\item[\hskip \labelsep {\bfseries #1}]}{\end{trivlist}}

\newcommand{\bd}{\begin{displaymath}}
\newcommand{\ed}{\end{displaymath}}
\newcommand{\be}{\begin{equation}}
\newcommand{\ee}{\end{equation}}
\newcommand{\bea}{\begin{eqnarray}}
\newcommand{\eea}{\end{eqnarray}}
\newcommand{\bda}{\begin{eqnarray*}}
\newcommand{\eda}{\end{eqnarray*}}
\newcommand{\ba}{\begin{array}}
\newcommand{\ea}{\end{array}}

\newcommand{\ra}{\rightarrow}

\newcommand{\lra}{\longrightarrow}

\newcommand{\dd}{\mbox{\rm\,d}}

\begin{document}

\title{\huge{ \textbf{Theoretical connections between mathematical neuronal models
 corresponding to different expressions of noise}}}

\author{Gr\'{e}gory Dumont\footnote{\'{E}cole Normale Sup\'{e}rieure,
Group for Neural Theory, Paris, France, email: gregory.dumont@ens.fr},
          Jacques Henry\footnote{INRIA team Carmen, INRIA Bordeaux Sud-Ouest, 33405 Talence cedex, France,
          email: Jacques.Henry@inria.fr}
       and
         Carmen Oana Tarniceriu\footnote{Intersdisciplinary Research Department - Field Sciences,
         Alexandru Ioan Cuza University of Ia\c{s}i, Romania
e-mail: tarniceriuoana@yahoo.co.uk}
      }


\date{}
\maketitle

\begin{abstract}
Identifying the right tools to express the stochastic aspects of neural activity has proven to be one of the biggest
challenges in computational neuroscience.  Even if there is no definitive answer to this issue, the most common procedure to express
this randomness is the use of stochastic models. In accordance with the origin of variability, the  sources of randomness are classified as
intrinsic or extrinsic and give rise to distinct mathematical frameworks to track down the dynamics of the cell.
While the external variability is  generally treated  by the use of a Wiener process in models such as the Integrate-and-Fire model,
the internal variability
is mostly expressed via a random firing process. In this paper, we investigate how those distinct expressions of
variability can be related. To do so, we examine the probability density functions to the corresponding stochastic models and
investigate in what way they can be mapped one to another via integral transforms. Our theoretical findings offer
a new insight view into the particular categories of variability and it confirms that,
despite their contrasting nature,
the mathematical formalization of internal and external variability are strikingly similar..
\end{abstract}

\section{Introduction}

\hspace{0.5cm}The presence of variability in the neural activity is well documented nowadays \cite{noiseNS}.
In vivo as well as in vitro experiments stand for the evidence of irregular behavior of neuronal activity. For instance, spike trains of individual cortical neurons in vivo are highly irregular \cite{ShN94,SK93}, and, apart from
randomness observed in spontaneous neuronal activity, recordings of in vitro experiments for input stimuli
without any temporal structure exhibited irregular behavior of neural activity \cite{gerstner}.
It is commonly accepted now that the random influences over the neuronal firing activity to be designated as noise \cite{andre01}.
Among the most notable
sources of noise, few are usually reminded : thermal noise \cite{manw},
the effect of signal transmission in a network \cite{manw}, randomness of excitatory and inhibitory connections \cite{BH01},
global network effects or the finite number of ionic channels on a neuronal patch \cite{white}.

According to the location where the noise is generated, it has become a common procedure to classify these
sources of noise as extrinsic or intrinsic \cite{gerstner}.
While  intrinsic noise usually refers to random features underlying
the firing process, therefore generated at cell level, the  extrinsic noise is generally attributed to the network effects
and signal transmission over the cell.

An important step toward the recent development in theoretical neuroscience consists in a deep understanding of the essence
of this variability. However, its mathematical formalization is still an open problem; it has been a subject of intense research,
(see for instance \cite{gerstner} for a discussion on this issue) and many recent papers are trying to suitably
mathematically model these effects.  A typical way to mathematically model random processes is by the use of stochastic differential
equations.  Nevertheless, the particular form of the noise terms and their incorporation into stochastic neuron models are still
subject of controversy.
Among others, two approaches gain more visibility in the last decade; each of them corresponds to different
 treatment of noise.

  The external influences over the transmembrane potential of a single neuron is usually modeled
in the form of a Gaussian white noise and gives rise to
an Ornstein Uhlenbeck process.
In the probability theory, the representation of a  random variable can be done by the use of the so called
 probability density function (pdf), which describes the likelihood for the stochastic process to take on a specific value.
Since a stochastic equation of Langevin type can be translated into a Fokker-Planck (FP) equation,
 this allows the representation of the external noise category as a diffusion process \cite{gardiner}.
 In particular, the FP equation \cite{gardiner} for an OU process  belongs to the
 most prominent models in the literature.

 It should be stressed that the use of pdf concept in the field of mathematical neuroscience has already a long history,
as it can be seen in \cite{rinzel,abbott}, and it has lead to revealing new insights into phenomena related to neuronal behaviors.
The resulted mathematical formalism is in particular pertinent for the simulation of large sparsely connected
populations of neurons in terms of a population density function \cite{sirovich,nykamp,us}.
 The formulation of the dynamics of neural networks in terms of population
densities made possible mathematical descriptions of phenomena that emerge at population level, such as oscillations and synchrony
caused by recurrent excitation  \cite{DH,DH01,carillo01}, by  delayed inhibition feedback \cite{BH01}, by both
recurrent excitation and inhibition \cite{B02}, and by gap junction \cite{brunel01}, the emergence of neural cascade
 \cite{cascade01,cascade02}, and of self criticality with synaptic adaptation \cite{critical}.
 For connections between models corresponding to probability respectively population densities, we refer to
\cite{nykamp,knight,K2000,BH01}.

To account for the intrinsic variability, neuronal models incorporating a stochastic mechanism of firing have been considered.
  In this framework, the state of the neuron follows a deterministic trajectory and each spike occurs with a certain probability
given in the form of an escape rate or stochastic intensity of firing. This assumption lead  to the introduction of
 models where  only the time passed since the last spike influences the dynamics.
 The associated pdf to such a model has the form of a
 so-called age-structured (AS) system \cite{gerstner}, \cite{perthame04}.
Such a process is a special form of renewal processes, which are a particular
category of point processes having the particularity that the times between two successive events are independent and
identically distributed. The main assumption upon which the process is built is that the whole history that happened before
the last event (firing time) is forgotten. It is therefore suitable to consider such a process for the case where neurons do not
show strong adaptation.

These different treatments of noise in the form of the two reminded models have been shown though to generate similar
statistics of spike trains \cite{gerstner}. An approximation method has been presented in \cite{plesser} to explain this
similarity. However, no exact relation between the solutions of this models has been yet proven. We did make a first step in this
direction in \cite{usdt} by mapping the solution to the AS system into the solution to the FP equation. The relation that we proved
is only partial since the reverse mapping of the solutions still lacked. To completely tackle this problem, we have investigated
therefore the possibility of giving an analytical transform of the AS system into the FP equation.
Our theoretical findings that we present here
highlight an unforeseen relationship between the FP
equation and the AS (of von-Foester-McKendrick type) system, which allows in
particular that qualitative properties of the solutions to be
transferred one-to-another.
 This explains not only the
similarities of the observed statistical activity of both models but
also help us rise a unitary vision over the different formalisms
used to describe different noise expressions acting over a neural cell.

The paper is structured as follows: We will start by reminding the two approaches of noise implementation and the main characteristics
of the corresponding models in the first section. As the kernels of our integral transforms depend on the solutions to the
forward and backward Chapman-Kolmogorov (CK) equation for a probability density function on an inter-spike interval, we will
roughly remind in the second section the derivation and the meaning of these two systems.
Our main results are given in the third section, and
we end this paper by some conclusive remarks and discussing possible extensions of the present work. Finally, all along the paper,
we present numerical simulations to illustrate the models presented in it.

\section{Two standard approaches in expressing neuronal variability}

Due to its stochastic nature, the anticipation of the evolution in time of the state of a neuron can not be made deterministically but
 only probabilistically. In computational neuroscience, a first step toward the description of neural variability was made in
 \cite{Stein67} and \cite{GM64}. In this section, we remind the reader two standard mathematical formalizations of neural
 variability in use nowadays. First we introduce the model that incorporates  an extrinsic source of noise in the form of the
 noisy leaky integrate-and-fire model,
 and then present its associated FP equation. Next, we turn to a model that expresses the intrinsic noise via a stochastic firing
 intensity which will lead to characterizing the evolution of the state of a neuron in the form of an AS system.

\subsection{Extrinsic noise. Fokker-Planck formalism}
\begin{figure}[]
\begin{center}
    \includegraphics[width=8cm]{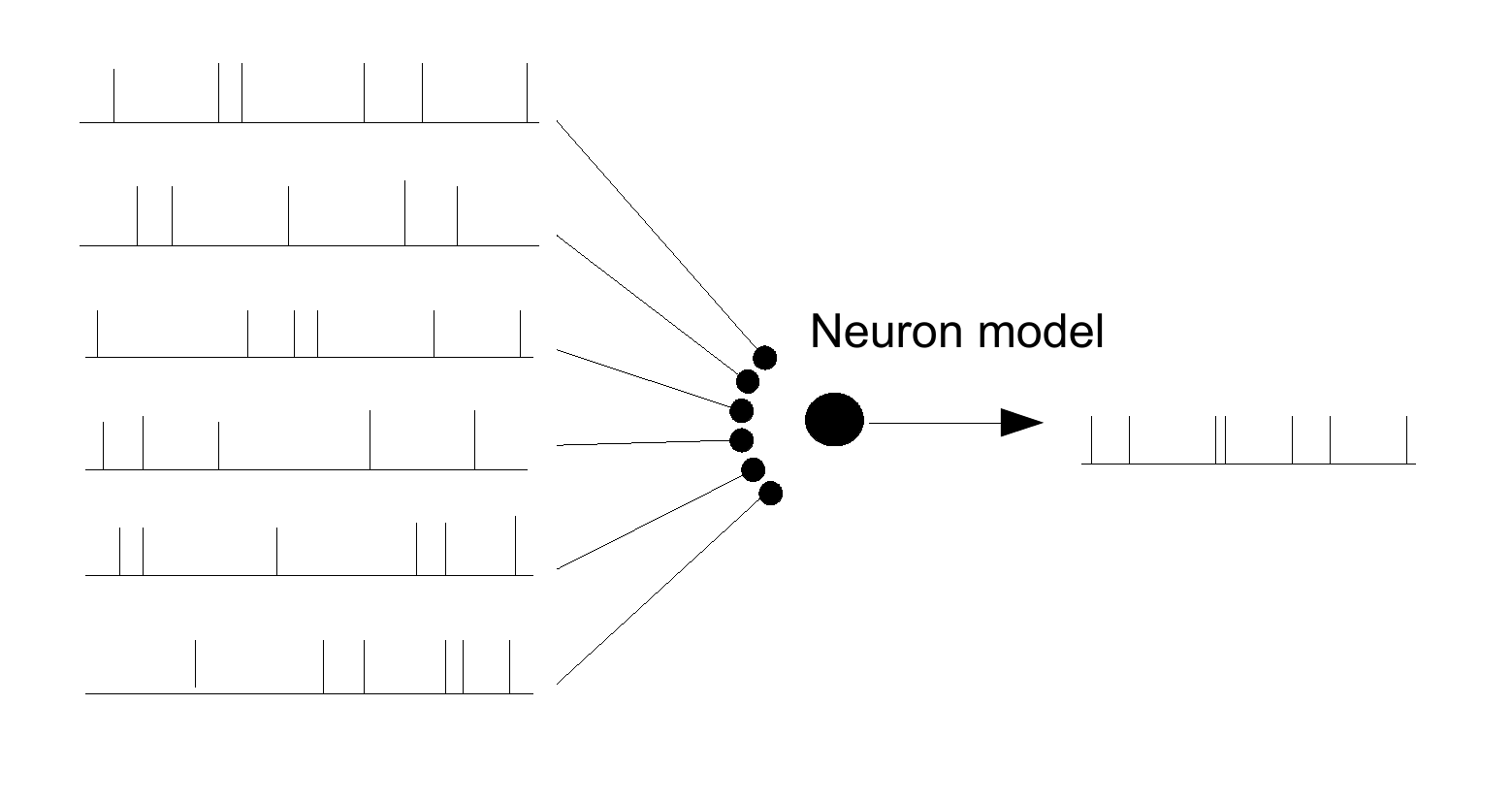}
  \caption{Schematic representation of a point process model. The model is referred as to be a point process since it takes
  into account only the time events and the complex
   shape of the action potential is not explicitly modeled. Point process models mostly focus on the
   input/ouput relationship, i.e. the relationship between the input spike trains the cell receives via synaptic afferent and its
   response to it. The model is
   said stochastic when the input or the firing process is random. }\label{Fig01}
      \end{center}
\end{figure}

As a common procedure to handle the extrinsic variability, noise terms are explicitly added to the mathematical models that describe
the evolution of the state of the nervous cell. In this way, the evolution of the state of a neuron is viewed as a random process
which is described by a stochastic differential equation (SDE). Throughout this paper, we will illustrate all the theoretical considerations
related to the extrinsic variability for the specific case of the noisy leaky integrate-and-fire (NLIF) model \cite{Izi}. The
integrate-and-fire model describes a point process (see Fig. \ref{Fig01}) and is largely used
because it combines a relative mathematical simplicity with the capacity
to realistically reproduce observed neuronal behaviors \cite{naud}. The model idealizes the neuron as a simple electrical circuit
consisting in a capacitor in parallel with a resistor driven by a given current and was first introduced by Lapique in 1907, see
\cite{B01,abbott01} for historical considerations and \cite{Lapique} for a recent English translation of  Lapique's original
 paper.

To be more specific, the model describes the subthreshold dynamics of a single neuron membrane's potential and a reset mechanism
to account for the onset of an action potential: A spike occurs whenever a given threshold $V_{T}$ is reached by the membrane
potential variable $V$. Once the firing event occurs, the membrane potential is right away reset to a given value $V_R$. In the
subthreshold regime, the membrane potential's dynamics is given by
\begin{equation*}
\tau \frac{d}{dt}V(t)=-g (V(t)-V_L)+\eta (t), \\
\end{equation*}%
where $V(t)$ is the membrane potential at time $t$, $\tau$ is the membrane capacitance, $g$ - the leak conductance, $V_L$ -
the reversal potential and $\eta (t)$ - a gaussian white noise, see \cite{Burkitt} for a recent review and see \cite{Izi} for other
spiking models. In what follows, we will use a normalized version of the above equation, i.e. we define $\mu$ as the bias current
and  $v$ the membrane's potential which will be given by

\begin{equation*}
\mu=   \frac{V_L}{V_T}, \quad v=\dfrac{V}{V_T}, \quad v_r=\frac{V_R}{V_T}.
\end{equation*}%
After re-scaling the time in units of the membrane constant $g/ \tau $, the normalized model reads

\begin{equation}
\label{IF}
\left\{
\begin{array}{l}%
\frac{d}{dt}v(t)=\mu-v(t)+\xi(t) \\
\text{If} \quad v>1 \quad \text{then} \quad v= v_r.
\end{array}%
\right.
\end{equation}%
 $\xi(t)$ is again a Gaussian white noise stochastic process with intensity $\sigma$:

\begin{figure}[]
\begin{center}
    \includegraphics[width=17cm]{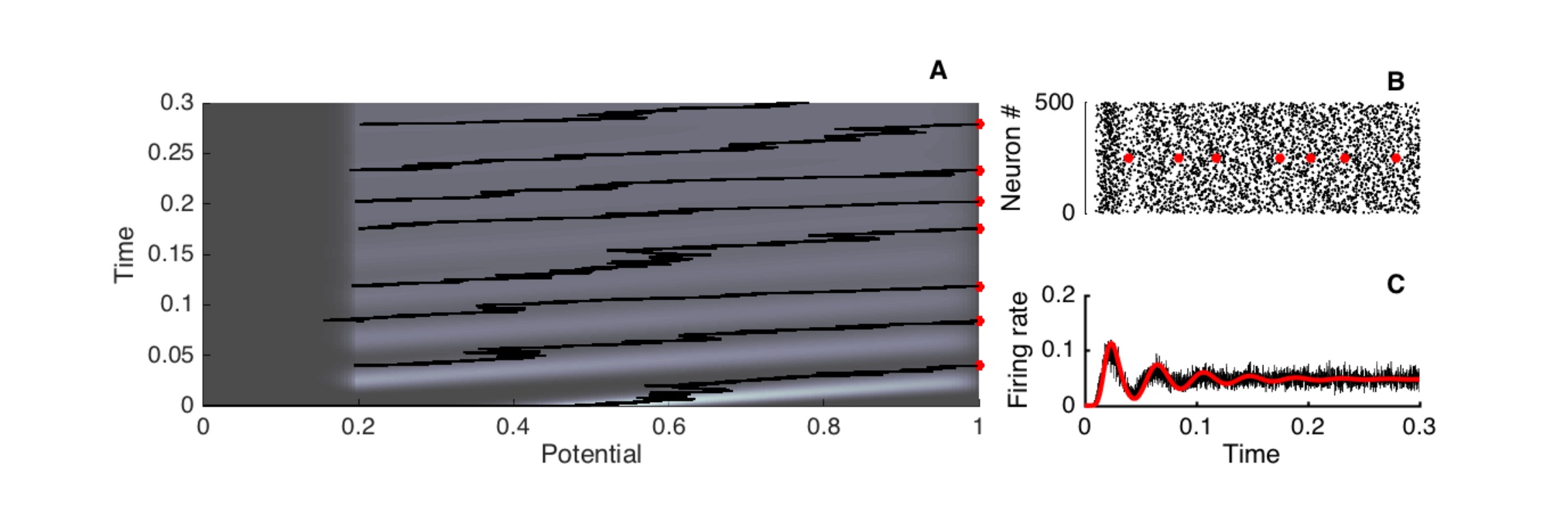}
   \caption{Simulation of the stochastic process (\ref{IF}) and of its associated FP equation (\ref{model}). A)
   Evolution in time of the density probability; the brightness is proportional to the probability density and
   the black line illustrates one realization of the process. B)
    Raster plot depicting the precise spiking time of the cell over different trials. The red dots correspond to the particular
     realization of panel A. C)
    The firing rate given by the FP equation, (\ref{firingP}), in red and by many realizations of the Langevin equation (\ref{IF}).  A gaussian was taken as initial condition; the parameters of the simulation are: $v_r=0.3$, $\mu=20$,
      $\sigma =0.4$.}\label{Fig02}
      \end{center}
\end{figure}

\begin{equation*}
 \langle \xi(t) \rangle =0, \quad  \langle \xi(t)\xi(t')  \rangle =\sigma \delta(t-t').
\end{equation*}%
The  NLIF model was introduced in \cite{knight72} and generalizations of it can be found
in the more recent work \cite{gerstner}. The first equation in (\ref{IF}) is a Langevin equation that contains a deterministic part
 expressed by the drift term
$\mu-v(t)$ and a stochastic part in the form of the noise term $\xi(t)$. The second line in (\ref{IF}) describes the onset
of an action potential and the reset mechanism.

One popular way to deal  with a SDE  is to write down the associated Fokker-Planck equation for the associated
probability density function (pdf).
 In the case of the NLIF model, the associated pdf
 $p(t,v)$ express the likelihood of finding the membrane potential at a given time $t$ in a value $v$.
 Starting with the SDE (\ref{IF}) of Langevin type,
 the interested reader will found in \cite{BN} a rigorous derivation of the associated Fokker-Planck equation.
 For our specific case the FP equation takes the following form:

\begin{equation}\label{model}
\frac{\partial }{\partial t}p(t,v)+\overbrace{\frac{\partial }{\partial v}[(\mu -v)p(t,v)]}^{\text{Drift}}
-\overbrace{ \frac{ \sigma ^2}{2}     \frac{\partial^2 }{\partial v^2}p(t,v) }^{\text{Diffusion}}  =
\overbrace{\delta (v-v_{r}) r(t)}^{\text{Reset}}.\quad
\end{equation}
The equation above expresses three different processes: a drift process due to the deterministic part in the NLIF model,
a diffusive part which is generated by the action of noise and a reset part which describes the re-injection of the neurons that
just fired into the reset value $v_r$. An absorbing boundary condition is imposed at the threshold value

\begin{figure}[]
\begin{center}
    \includegraphics[width=17cm]{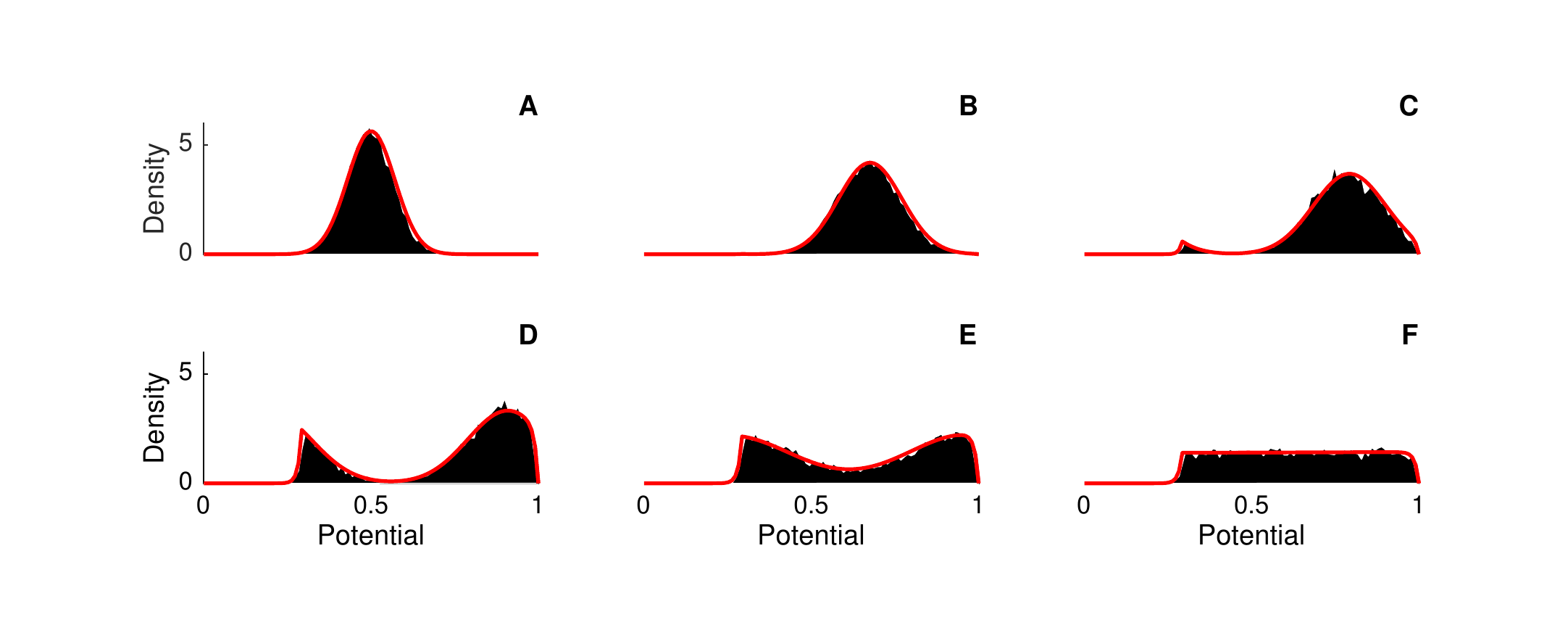}
   \caption{Simulation of the stochastic process (\ref{IF}) and of its associated FP system (\ref{model})-(\ref{firingP}).  A gaussian was taken
   as initial condition; the parameters of the simulation are:
   $v_r=0.3$, $\mu=20$, $\sigma =0.4$. The plots show the evolution in time of the solution respectively $t=0$, $t=0.1$,
    $t=0.3$, $t=0.5$,  $t=0.7$, $t=7$ for the respective panel A-B-C-D-E-F.  }\label{Fig03}
      \end{center}
\end{figure}

\begin{equation}
p(t,1)=0,
\end{equation}
which expresses the firing process,
and a reflecting boundary condition at $v=-\infty$
\begin{equation}
\lim\limits_{v \to -\infty} (-\mu +v)p(t,v) +\frac{ \sigma ^2}{2}     \frac{\partial }{\partial v}p(t,v) =0,
\end{equation}
which states that there is no flux passing through this boundary.

 The firing rate $r(t)$ is defined as the flux at the threshold $v=1$:
 \begin{equation}
\label{firingP}
 r(t)  = -\frac{ \sigma ^2}{2}     \frac{\partial }{\partial v}p(t,1).
\end{equation}
To uniquely determine a solution, an initial condition is given:
\be\label{IC}
p(0,v)=p_0(v).
\ee
Using the boundary conditions and the expression of $r(t)$ given by (\ref{firingP}), one can easily check the conservation
property of the equation (\ref{model}) by  directly integrating it on the interval $(-\infty,1)$, so that,
if the initial condition  satisfies

\begin{equation}
\int  _{-\infty}^{1} p_0(v)\,dv=1,
\end{equation}
then the solution to (\ref{model})-(\ref{IC}) necessarily satisfies the normalization condition

\begin{equation}
\label{normalisation}
\int  _{-\infty}^{1} p(t,v)\,dv=1.
\end{equation}

We present in  Fig \ref{Fig02} a simulation of the FP model (\ref{model})-(\ref{IC}). The numerical results illustrate the stochastic
process (\ref{IF}) and the time evolution of its associated probability density. In  Fig \ref{Fig02}, the time evolution of the
density is depicted in the first panel while the black line only gives one particular realization of the Langevin type equation
(\ref{IF}). Note the effect of noise on the dynamics of the membrane potential: For this particular neuron, the firing events occur
whenever the membrane potential reaches the threshold, which is highlighted by the presence of a small red dot in our simulation.
The firing activity is also represented, the red curve corresponding to the FP equation (\ref{model})-(\ref{IC}) and the blue curve
to the stochastic process (\ref{IF}).

To get a better understanding of the time evolution of this probability density, we present in Fig. \ref{Fig03} different snapshots
of the simulation. Under the drift and the diffusion effects, the density function gives a non zero flux at the threshold, and this flux
is reset to $v_r$ according to the reset process. This effect can be seen clearly in the third panel of the simulation presented in
Fig. \ref{Fig03}.
Asymptotically, the solution  reaches a stationary density, which is shown in the last panel of Fig. \ref{Fig03}.

\subsection{Intrinsic noise. McKendrick-von Foerster formalism}
To account for the intrinsic variability and the randomness in the firing process, a typical procedure is
to consider deterministic trajectories and to assume that a firing
event can occur at any time according to a certain probability rate \cite{G95}.
This rate is often called escape rate or stochastic
intensity of firing \cite{gerstner} and will be denoted by $S$ throughout this paper. Therefore the implementation of intrinsic noise
lets the deterministic trajectories, expressed by one of the known single neuron models, unaffected, but instead influences the firing
time, which is no longer deterministic. In this setting, a neuron may fire even though the formal threshold has not been reached yet
or may stay quiescent after the formal threshold has been crossed. Usually the expression of $S$ depends on the momentary distance of
the membrane potential to the formal threshold of firing, and therefore the process can be fully characterized by the amount
of time passed by since the last action potential.

A random variable described in this way is known as a renewal process which is a class of stochastic processes characterized by the
fact that the whole history of events that had occurred before the last one can be forgotten \cite{cox}. Note that the use of the renewal
theory in neuroscience is justified by the fact that the neuron is not supposed to exhibit strong adaptation,
see \cite{DSNG} for the case of
adapting neurons. In \cite{gerstner}, \cite{plesser} there is treated the case of time dependent input which leads to the use of
the more general case of non-stationary renewal systems. We will restrict in this paper to the stationary renewal theory since we only
consider the case of time-independent stimulus, and therefore, the only variable on which the escape rate will depend is the age $a$,
which stands for the time elapsed since the last spike. In such a framework, the neuron follows a stochastic process where the
probability of surviving up to a certain age, $\mathcal{P}\left( a\right)$, is given by

\begin{figure}[]
\begin{center}
    \includegraphics[width=17cm]{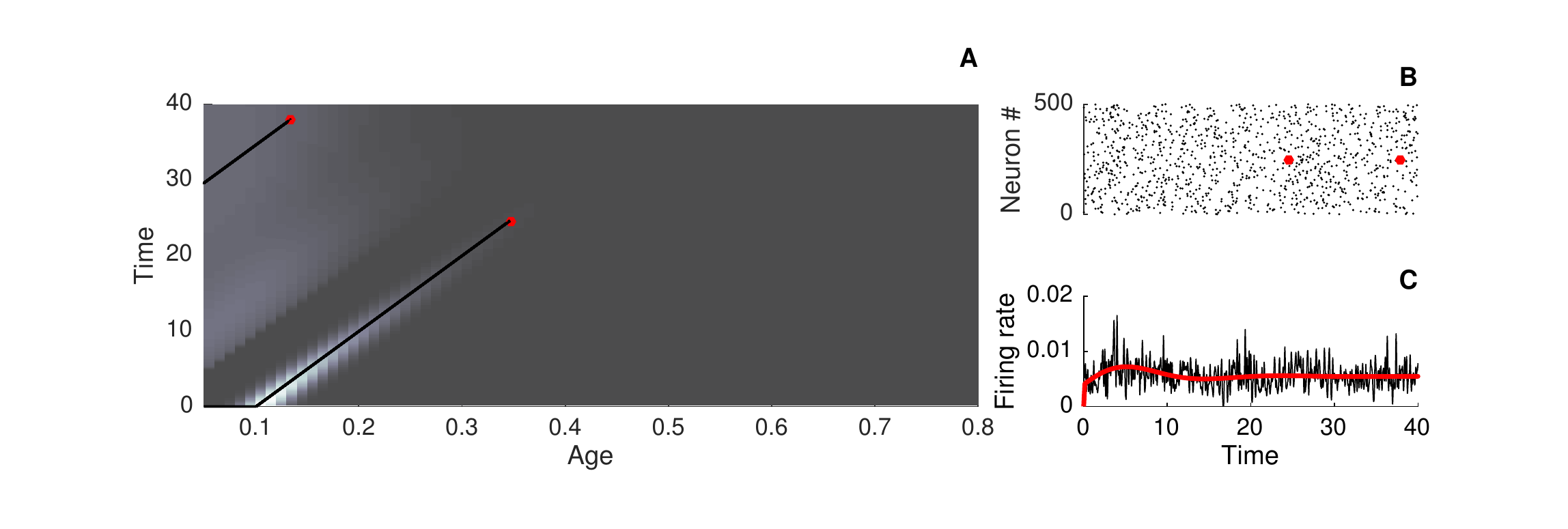}
   \caption{Simulation of the stochastic process (\ref{SA}) as well as of its associated AS equation (\ref{age})-(\ref{id}).
    A)
   Evolution in time of the density probability; the brightness is proportional to the probability density and the black line  illustrates one realization of the process. B)
    Raster plot depicting the precise spiking time of the cell over different trials. The red dots correspond to the particular
     realization of panel A.  C)
    The firing rate given by the AS equation (\ref{age}) in red and by many realizations of (\ref{SA}).
    A gaussian was taken as initial condition. The function $S$ used in the simulations is given by (\ref{SF})
    with parameters $v_r=0.3$, $\mu=20$, $\sigma=0.4$. }\label{Fig04}.
      \end{center}
\end{figure}

\begin{equation}\label{SA}
\mathcal{P}\left( a\right)= \exp\{-\int_0^a S(s) \dd s\},
\end{equation}%
where the specific choice of $S$ determines the properties of the renewal process. We will come back with  details about the
form of $S(a)$ in the following sections.

To better understand the renewal process, one can focus on the probability density function that describes the relative likelihood
for this random variable to take on a given value. This description leads to a so-called age-structured
 system (\cite{gerstner}) which consists in a partial differential equation with non-local boundary condition that is
 famous in the field of population dynamics, and which, in our specific case, is in the form of
 the McKendrick-von Foerster model \cite{vonF}. Denoting by $n(t,a)$ the
probability density for a neuron to have at time $t$ the age $a$, then the evolution of $n$ is given by

\begin{figure}[]
\begin{center}
    \includegraphics[width=17cm]{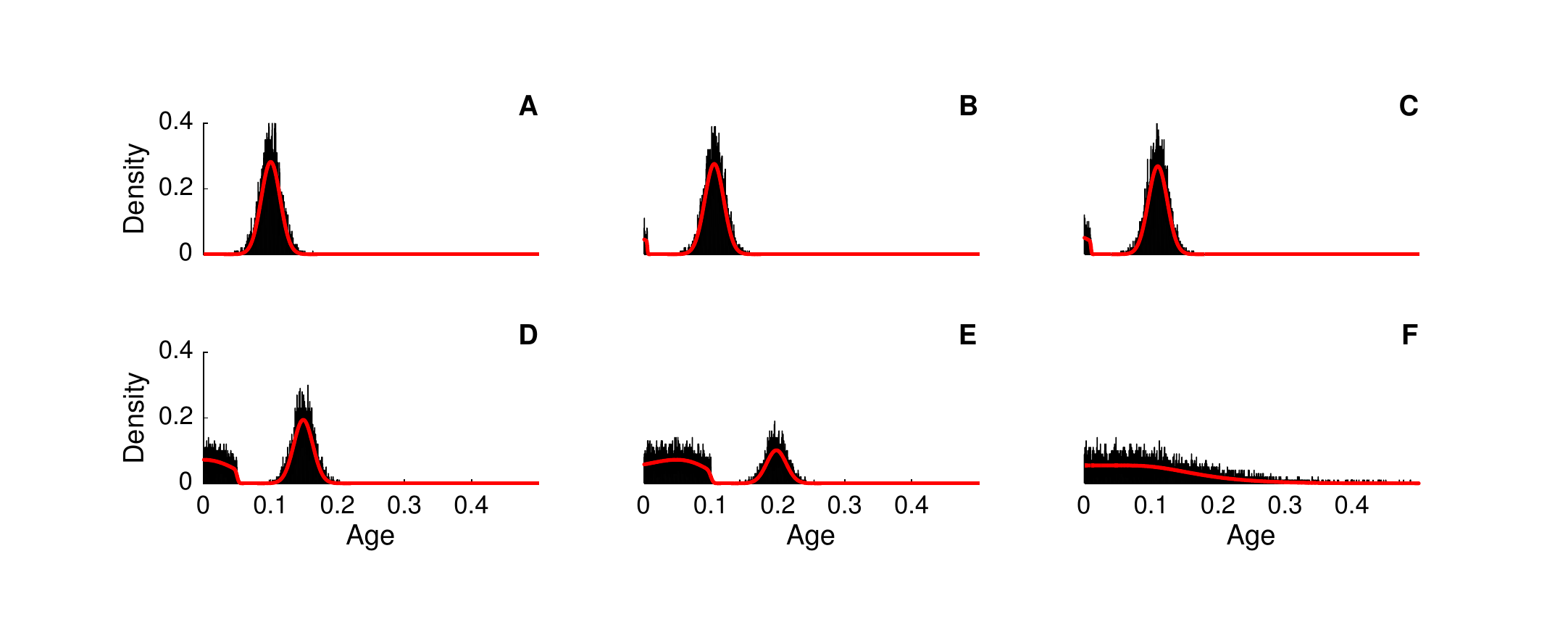}
   \caption{Simulation of the stochastic process (\ref{SA}) as well as its associated AS equation (\ref{age})-(\ref{id}).
   The different panels A-B-C-D-E-F correspond to different times of the simulation.
   The function $S$ used in the simulations is given by (\ref{SF})
    with the same parameters as in Fig. \ref{Fig04}.  }\label{Fig05}
      \end{center}
\end{figure}

\begin{equation}\label{age}
\frac{\partial }{\partial t}n(t,a)+ \overbrace{\frac{\partial }{\partial a}n(t,a)}^{\text{Drift part}}+
\overbrace{S(a)n(t,a)}^{\text{Spiking term}}=0.\\
\end{equation}%
Because once a neuron triggers a spike  its  age is reset to zero, the natural boundary condition to be considered is

\be
\overbrace{n(t,0)=r(t)}^{\text{Reset}},\quad t>0,
\ee
where $r(t)$ is the firing rate and is given by

\begin{equation}
\label{firing}
r(t)= \int_{0}^{+\infty}S(a)n(t,a) \,da,\quad   t\geq 0.
\end{equation}
To completely describe the dynamics of $n$, an initial distribution is assumed known:

\be\label{id}
n(0,a)=n_0(a), \quad  a>0.
\ee
Using the boundary condition and the expression of $r(t)$ given by (\ref{firing}), one can check easily the
conservation property of the equation (\ref{model}) by  integrating it on the interval $(0,\infty)$, so that if the initial
 condition  satisfies

\begin{equation*}
\int  _{0}^{+\infty} n_0(a)\,da=1,
\end{equation*}
the solution at any $t>0$ satisfies the normalization condition

\begin{equation}
\label{normalisation}
\int  _{0}^{+\infty} n(t,a)\,da=1.
\end{equation}

We present in Fig. \ref{Fig04} a simulation of the escape rate model and of its associated probability density function $n$.
Note that the main distinction on the stochastic aspect is that the noise does not act on the trajectory but only on the initiation
of action potential. Again, we have made a comparison between the stochastic process (black curve) and the evolution of the density
function (red curve).
The simulation starts with a Gaussian as initial condition (the first panel of Fig. \ref{Fig05}). Under the influence of the drift
term, the density function advances in age, which is clearly seen in the upper plots of Fig. \ref{Fig05}. After the spiking process,
the age of the neuron is reset to zero. The effect is well perceived in the lower panels of Fig. \ref{Fig05}. As expected from the
model, the density function converges to an equilibrium state.

\section{Chapman-Kolmogorov equation}
The models introduced above, (\ref{model})-(\ref{IC}) respectively (\ref{age})-(\ref{id}),
have been shown to exhibit the same statistical activity \cite{gerstner}, even though the stochastic processes
that they illustrate are conceptually different. To explain this behavior, we have been looking in which way the solutions of the
two models can be analytically related.

The integral transforms that will be defined later on and which give this connection make use of the solutions to backward and
forward Chapman-Kolmogorov (CK) equations for the stochastic process described by the NLIF model defined on an inter-spike interval.
Before tackling this  issue, we will first remind few theoretical considerations about the CK equation, how the forward
and backward systems are derived from it and the interpretations of the solutions to both systems. For a complete derivation
and analysis of it, we refer to the classic work \cite{gardiner}, but we also refer to \cite{BN} for discussions about the forward
and backward CK equations in biological processes.

\subsection{Markov property}
The  Chapman-Kolmogorov equation is formulated in terms of conditional probabilities and it is built on the main assumption that the
stochastic process in question satisfies the Markov property:
The future states of the system depend only on the present state.
Defining the conditional probability $\mathfrak{p}(t,v|s,w)$ as the probability to be in the state $v$ at time $t$
given that it started at time $s$ from $w$, the CK equation reads:
\begin{equation}\label{CK}
\mathfrak{p} (t,v|s,w)=\int \mathfrak{p}(t,v|t',v')\mathfrak{p}(t',v'|s,w)\dd v'.
\end{equation}
Roughly speaking, the CK equation simply says that, for a Markov process, the transition from $(s,w)$ to $(t,v)$ is made in two steps:
 first the system moves from $w $ to $v'$ at an intermediate time $t'$ and then from $v'$ to $v$ at time $t$; the transition
 probability between the two states is calculated next by integrating over all possible intermediate states.

Starting from the integral CK equation (\ref{CK}), the forward CK equation, also known as the FP equation, as
well as the backward CK equation are derived under the assumption of  sample paths continuity  of the stochastic
process (\cite{gardiner}). While the FP equation is obtained by considering the evolution with respect to present time $t$,
the backward CK equation is derived when considering the time development of the same conditional probability with respect to
initial times.


To keep a certain level of simplicity, we shall not refer in this paper to the general form of both equations,
and we refer to \cite{gardiner} for the general case as well as for the derivation and interpretations in particular cases. We remind that, since we have considered in this paper the constant stimulus case, we deal with a stationary stochastic process.
A stationary process has by definition the same statistics for any time translations, which implies that the associated
joint probabilities densities are time invariant too; due to the relation between the conditional and joint probabilities
and also to the Markov property, this allows expressing the conditional probabilities in the CK equation in terms of time differences.
In particular, the FP equation as well as the backward CK equations can be written as in the following.

\subsection{Forward and backward Chapman-Kolmogorov equations}
We shall consider first the conditional probability for a neuron that started at the last firing time $s$ from the corresponding
state $v_r$ to be at time $t$ in a state $v<1$; since in the following we shall keep the initial variables fix, and use the
property of stationary processes, we can write down the associated FP equation in terms of a joint probability.
We deal therefore with the following FP equation for the probability density function
$$\varphi(a,v):=\mathfrak{p}(t-s,v|0,v_r)$$
 for a neuron to have
at age $a:=t-s$  the membrane's potential value $v$, where we have skipped from the notation the variables which are kept fix:

\be\label{FPi}
\frac{\partial}{\partial a}\varphi(a,v)+\frac{\partial}{\partial v}[(\mu-v)\varphi(a,v)]
-\frac{\sigma^2}{2}\frac{\partial^2}{\partial v^2}\varphi(a,v)=0.
\ee
The initial condition

\be
\varphi(0,v)=\delta(v-v_r)
\ee
is natural since the probability density for a neuron that started at $a=0$ from the potential $v_r$ to have at
the same age $a=0$ a potential $v$ can only be $\delta(v-v_r)$.

Also, due to the firing process at threshold value $v=1$, an absorbing boundary condition is imposed:

\be
\varphi(a,1)=0.
\ee
At $v=-\infty$, it is required that the flux should be zero, which corresponds to a reflecting boundary:

\be\label{bcf}
\lim_{v\ra-\infty}\left[(v-\mu)\varphi(a,v)+\frac{\sigma^2}{2}\frac{\partial}{\partial v}\varphi(a,v)\right]=0.
\ee
 Note that, in contrast to the model (\ref{model})-(\ref{IC}), here there is no conservation property
 required; it is suitable therefore to think of $\varphi$ as a probability density on an inter-spike
 interval, therefore the reset term in (\ref{model}) does not appear in this case. Nevertheless, the flux
 at the threshold $v=1$  is a quantity of big interest since it gives
 {\it the inter-spike interval distribution (ISI)} function, that was introduced generically in the previous section, for a neuron that
 started at age $a=0$ from the reset value $v_r$

  \begin{figure}[]
\begin{center}
    \includegraphics[width=17cm]{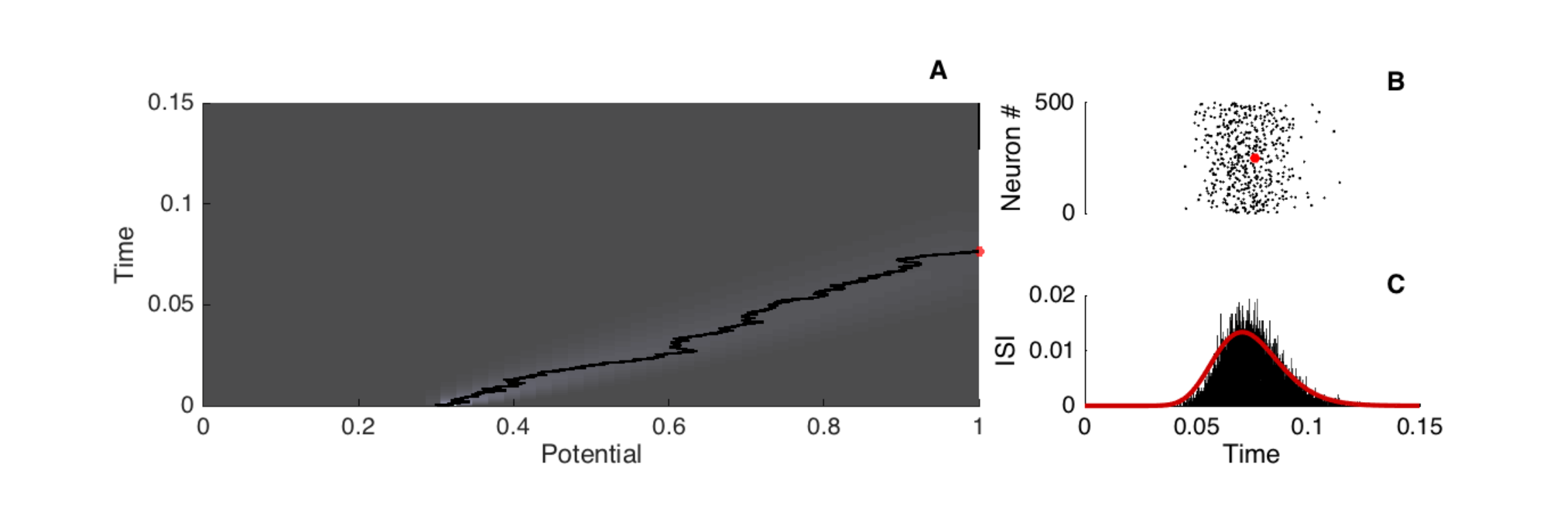}
   \caption{Simulation of the stochastic process as well as its associated FP equation (\ref{FPi}). A)
   Evolution in time of the density probability;the brightness is proportional to the probability density and the black line illustrates one realization of the process. B)
    Raster plot depicting the precise spiking time of the cell over different trials. The red dot corresponds to the particular
     realization of panel A. C)
    The ISI obtained from the FP equation (\ref{FPi}) in red, and by many realizations of the Langevin equation (\ref{IF}).
    The parameters of the simulation are: $v_r=0.3$, $\mu=20$,
      $\sigma =0.4$.  }\label{Fig06}
      \end{center}
\end{figure}

 \be\label{isi}
 ISI(a)=-\frac{\sigma^2}{2}\frac{\partial}{\partial v}\varphi(a,1).
 \ee
The central notion of the renewal theory is the {\it interval distribution} of the events; it has the role of
predicting the probability
of the next event to occur in the immediate time interval. In neuroscience context, the popularized notion is the
inter-spike interval distribution.
There is a tight connection between the ISI distribution and the survivor function $\mathcal{P}$ due to their interpretations.
 Starting with (\ref{isi}), one can define next  the survivor function $\mathcal{P}$ as  the probability to survive up to age $a$
 for a neuron that started at age $a=0$ from the position $v_r$ as
 \be
 \mathcal{P}(a)=\int_{-\infty}^1 \varphi(a,v)\dd v.
 \ee
 Note that the  relation reminded above that takes place between these two functions, i.e.

\begin{equation}\label{ISI}
ISI(a)=-\frac{\partial}{\partial a}\int_{-\infty}^{1} \varphi(a,v) \dd v,
\end{equation}
is verified  by integrating (\ref{FPi}) with respect to $v$ on the whole potential values interval and using the
boundary conditions.
Also, one may check directly that

\begin{equation*}
\int_{-\infty}^{1} \varphi(a,v) \,dv=1-\int_0^a ISI(s)\dd s.
\end{equation*}
Both these probabilities are defined here as they were introduced in \cite{gerstner}.

In the case of backward CK equation, the conditional probability for a neuron that started at age $a=0$ from a potential
value $v$ to survive up to the age a is considered,
$$\psi(a,v):=\mathfrak{p}(t-s,w|0,v),$$
where, again, we have skipped from the notation the variables which are kept fix.

 We did use the same name for the time variable in the definition of $\psi$, i.e. age, for the sake of
fluency, although inhere $a$
represents simply the time passed since the initial time considered, $a=t-s$, and
not the time passed since the last spike.
Note that this probability is the same with the probability that at age $a$ the present state $w$ for a neuron that
started at $a=0$ from $v$ did not reach yet the threshold $v=1$. With this respect, the backward CK equation gives the evolution
with respect to initial states:

\be\label{FirstPassage}
\frac{\partial}{\partial a}\psi(a,v)-(\mu-v)\frac{\partial}{\partial v}\psi(a,v)
-\frac{\sigma^2}{2}\frac{\partial^2}{\partial v^2}\psi(a,v)=0.
\ee
 Since, obviously, the probability to survive without firing at age $a=0$ for a neuron that
starts at the same age from a potential value $v<1$ is equal to 1, the suitable initial
condition is

\begin{equation}
\psi(0,v)=1,
\end{equation}
and, due to the same interpretation, a neuron that reaches the threshold value $v=1$ at age $a$ has the probability of survival

\begin{equation}
\psi(a,1)=0.
\end{equation}
We shall impose at the boundary $v=-\infty$ the condition

\be\label{bcb}
\lim_{v\ra-\infty}\frac{\partial\psi(a,v)}{\partial v}=0.
\ee
This last condition is obtained due to the corresponding boundary condition to the FP equation (\ref{FPi})
and to the duality of the functions $\varphi$ and $\psi$ \cite{gardiner}.

Note that, in contrast with the survivor function introduced above as in
 \cite{plesser}, $\psi$  expresses the probability of survival for a neuron that started at age
$a=0$ from a potential value $v<1$; then, obviously, the following relation must hold:
\begin{equation}\label{eql}
\int_{-\infty}^{1} \varphi(a,w) \,dw=\psi(a,v_r),
\end{equation}
equality that checks out immediately due to the definitions of both functions
as solutions to equations (\ref{FPi}) and (\ref{FirstPassage}) and using
 integration by parts:
\begin{equation*}
0=\int_0^a\int_{-\infty}^1\frac{\partial\psi(t,v)}{\partial t}\varphi(a-t,v) dv dt-\int_0^a\int_{-\infty}^1\psi(t,v)\frac{\partial \varphi}{\partial a}(a-t,v) dv dt=\psi(a,v_r)-\int_{-\infty}^1\varphi(a,v) dv
\end{equation*}

\begin{figure}[]
\begin{center}
    \includegraphics[width=11cm]{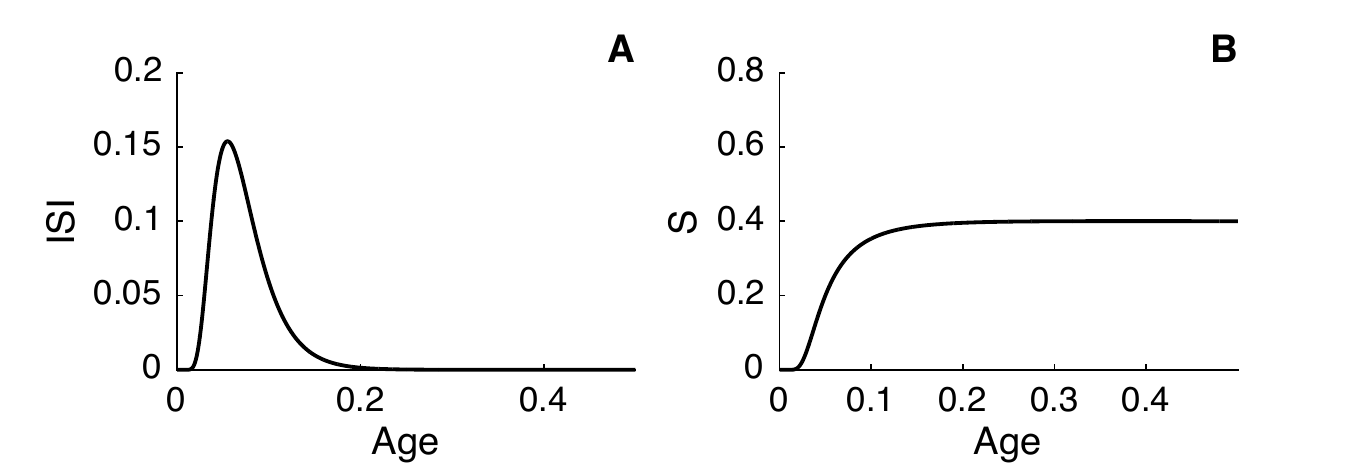}
   \caption{Illustration of the $ISI$ and the spiking rate function $S$. This illustration is obtained via a numerical simulation
   of the function $S(a)$ given by (\ref{SF}) in the panel A and its corresponding
   $ISI$ given by (\ref{ISI}) in the B panel.
    The parameters of the simulation are $v_r=0.7$, $\mu=5$, $\sigma =0.1$}\label{Fig08}
      \end{center}
\end{figure}

We now have all the necessary elements to properly define the age-dependent death rate corresponding to
the model (\ref{age})-(\ref{id}) as the rate of decay of the survivor function
\begin{equation}\label{SF}
S(a)=-\frac{ \frac{\partial}{\partial a}  \int_{-\infty}^{1} \varphi(a,w) \,dw}{\int_{-\infty}^{1} \varphi(a,w) \,dw}=
\frac{ISI(a)}{1 -\int_0^a ISI(s)\,ds }
\end{equation}
which has the interpretation that, in order to emit a spike, the neuron has to stay quiescent in the
interval $(0, a)$ and then fire at age $a$, see \cite{gerstner}. In Fig. \ref{Fig08}, numerical simulations of the age-dependent
death rate is presented. Let us notice that $S$ clearly defines  a positive function that converges towards a constant.

\section{Analytical links between the FP and AS systems}
Having set the theoretical framework, we are now ready to introduce our results regarding the connections between the two
systems that correspond to different mathematical treatments of variability. More precisely,
we will define  integral transforms that map one-to-another the solutions to the two models considered, i.e. (\ref{model})-(\ref{IC})
respectively
(\ref{age})-(\ref{id}). One single important
assumption over the initial probability densities of the systems is necessary for obtaining the result in
Theorem \ref{ITransformation}; that is:

\be\label{ir}
p_0(v)=\int_0^\infty \frac{\varphi(a,v)}{\int_{-\infty}^{1} \varphi(a,w) \,dw}n_0(a)\dd a.
\ee
As we stressed before, the FP equation does not give any information about the firing times, merely it can indicate the
condition for which a spike is triggered. Our relation simply assumes  that the repartition of the potentials at a given
initial time corresponds to an initial probability density of ages. It is assumed in this way that
the neuron has fired at least once.

\subsection{From diffusive noise to escape rate.}

In  \cite{usdt}, we have already studied the analytical transformation of the solution to the
model expressing the internal variability into the solution corresponding to the external one. Namely,
we have shown  that, once the relation (\ref{ir}) takes place, then for all times $t>0$, a similar relation between
the density of potentials respectively ages takes place. The result states:

\begin{theorem}\label{ITransformation}
Let $p$ a solution to (\ref{model})-(\ref{IC}) and $n$ a solution to (\ref{age})-(\ref{id}),
and $p_0(v)$ and $n_0(a)$ two corresponding
initial densities, respectively.
Then if $p_0$ and $n_0$  satisfy
\begin{equation}\label{C17060}
p_0(v)=  \int_{0}^{+\infty}  \dfrac{\varphi (a,v) }{ \int_{-\infty}^1  \varphi (a,w)   \, dw } n_0(a)  \, da,
\end{equation}
the following relation holds true for any time $t>0$:

\begin{equation}
\label{C1706}
p(t,v)= \int_{0}^{+\infty}\frac{\varphi (a,v) }{\int_{-\infty}^{1} \varphi(a,w) \,dw} n(t,a) \,da.
\end{equation}%
Here, $\varphi (a,v)$ is the solution to (\ref{FPi})-(\ref{bcf}).
\end{theorem}

In our above quoted work, the exact meaning of the solutions involved in (\ref{C1706})
and, consequently, of the integral (\ref{C1706}) is given. A similar relation between the respective
steady states of the models (\ref{model})-(\ref{IC}), (\ref{age})-(\ref{id}) has been also proven.

Note that, the probabilistic meaning of the integral transform given in Theorem \ref{ITransformation} can be interpreted
using Bayes' rule. Since $\varphi (a,v)$ is the probability density for
a neuron to be at  age $a$ and at potential $v$, the kernel of the transform can be interpreted as the probability density
for a neuron to be at potential value $v$ given that it survived up to   age $a$. Then, the product of this kernel with the
 solution $n(t,a)$,
which denotes the probability density  at time $t$ in state $a$, integrated over the all possible states $a$,
 gives indeed the probability density to be at time $t$ in the state $v$.

For more details about the result we refer to \cite{usdt}.

\subsection{From escape rate to diffusive noise}
Transforming the solution to the FP model into the solution to the
AS system is a little bit trickier. Before stating our result,
 let us  make few comments about the features of this problem. Note first that the very
nature of the age $a$ contains all the information about time that is needed to properly define the integral transform (\ref{C1706}).
On the contrary, to define an inverse transform, one faces the problem of having a kernel that must depend on time. A second important
aspect about the nature of
the AS formalism is that the variable $a$ also entails information about the last firing moment. Indeed, attributing an age to a
neuron presupposes that the considered neuron has already initiated an action potential.
From our perspective, the membrane potential variable $v$ does not carry out such information.

The main result of the paper is the following:

\begin{theorem}\label{Inverse}
Let $p$ be a solution to (\ref{model})-(\ref{IC}) and $n$ be a solution to (\ref{age})-(\ref{id}).

Then, for any $t$ such that $0\leq a < t$, the following relation holds true:

\begin{equation} \label{Inverse01}
n(t,a)=-\frac{\partial}{\partial a} \int_{-\infty}^1 \psi(a,v)p(t-a,v) \dd v.
\end{equation}

Moreover, if the compatibility condition between the initial distributions of potentials and ages
(\ref{C17060})
is assumed, then, for any $a\geq t$, a global relation between the corresponding probabilities at time $t$ takes place:
\begin{equation} \label{Inverse02}
\int_{t}^{+\infty} n(t,a) \,da  =  \int_{-\infty}^1\psi(t,v)p_0(v) dv.
\end{equation}
Here, $\psi(\cdot,v)$ the solution of the dual problem
(\ref{FirstPassage})-(\ref{bcb}).\\
\end{theorem}

\begin{remark}
All the solutions in Theorem \ref{Inverse} are taken in distributional sense. In this paper, we  intend
to maintain a certain level of simplicity by not going into details about the functional spaces in which the
solutions are considered. We do refer for  rigorous definitions of the solutions to the system (\ref{model})-(\ref{IC})
 to \cite{carillo01}. Also, a more general case of the system (\ref{age})-(\ref{id}) has been considered  in
\cite{perthame03}. In what follows, therefore, we preferred to keep the computations at a formal level so that to
preserve the intuitional meaning of the functions involved.
\end{remark}

\begin{remark}
Note that relation (\ref{Inverse02}) does only provide a global relation between $n(t,a)$ for $a>t$ and
$p_0(v)$, but not an exact expression of $n(t,a)$ as a function of $p_0(v)$; the explicit form of $n(t,a)$ is known
only if $n_0(a)$ is considered known.
\end{remark}

\begin{remark}
The only relation needed to define the transform $p\mapsto n$ for the whole interval $t>0$
is (\ref{Inverse01}) if one assumes that for any $a>t$,
in the time interval $[t-a,0]$  the FP system evolved in a stationary
regime with the same parameters as for $t>0$.
\end{remark}

\begin{proof}
We start our proof by noting that the explicit solution of the AS system can be calculated and is given by
\be
n(t,a)=\left\{ \begin{array}{ll}
                \frac{ \mathcal{P}(a)}{\mathcal{P}(a-t)}n_0(a-t),\quad t\leq a,\\
                \mathcal{P}(a)r(t-a),\quad t>a.
                \end{array}\right.
\ee
Let us consider next, for any $t>a$, the integral
\bd
\int_{-\infty}^1 \psi (a,v) p(t-a, v) \, dv.
\ed
A straightforward computation gives:
\begin{equation*}
 \frac{\partial }{\partial a}\int_{-\infty}^1 \psi (a,v) p(t-a, v) \dd v
 =  \int_{-\infty}^1  p(t-a, v) \frac{\partial }{\partial a} \psi (a,v) \dd v
 +\int_{-\infty}^1 \psi (a,v) \frac{\partial }{\partial a}  p(t-a, v) \dd v .
\end{equation*}
Due to the fact that $\psi$ is solution to (\ref{FirstPassage}), the first term in the right hand side can be written as
\begin{equation*}
\begin{split}
  \int_{-\infty}^1  p(t-a, v) \frac{\partial }{\partial a} \psi (a,v) \, dv = \int_{-\infty}^1  p(t-a, v)
  \left(  (\mu -v) \frac{\partial }{\partial v}\psi(a,v)+\frac{ \sigma ^2}{2}\frac{\partial^2 }{\partial v^2}\psi(a,v) \right) \dd v .
 \end{split}
\end{equation*}
Applying an integration by parts, we get
\begin{equation*}
\begin{split}
 & \int_{-\infty}^1  p(t-a, v) \frac{\partial }{\partial a} \psi (a,v) \dd v =
   \left[ p(t-a,v)\frac{\partial }{\partial v}  \psi (a,v) \right]_{-\infty}^1
  -\int_{-\infty}^1
 \frac{ \sigma ^2}{2}     \frac{\partial}{\partial v} \psi (a,v) \frac{\partial }{\partial v}p(t-a,v) \dd v   \\
 &   +\left[ (\mu -v)p(t-a,v)  \psi (a,v) \right]_{-\infty}^1
 -\int_{-\infty}^1\psi (a,v) \frac{\partial }{\partial v}[(\mu -v)p(t-a,v)]  \dd v  ,
 \end{split}
\end{equation*}
 and using the boundary conditions of both problems in $p$ and $\psi$, the last expression gives
\begin{equation*}
\begin{split}
 \int_{-\infty}^1  p(t-a, v) \frac{\partial }{\partial a} \psi (a,v) \dd v
 =- \left[ \psi (a,v)\frac{\partial }{\partial a} p(t-a,v)\right]_{-\infty}^1\\
 +\int_{-\infty}^1 \psi (a,v) \left(
 \frac{ \sigma ^2}{2}     \frac{\partial^2 }{\partial v^2}p(t-a,v)   -\frac{\partial }{\partial v}[(\mu -v)p(t-a,v)]  \right) \dd v.
 \end{split}
\end{equation*}
Using again the absorbing boundary conditions for $p$ and $\psi$, we finally get that the first term is indeed
\begin{equation*}
\begin{split}
  \int_{-\infty}^1  p(t-a, v) \frac{\partial }{\partial a} \psi (a,v) \dd v = \int_{-\infty}^1 \psi (a,v) \left(
 \frac{ \sigma ^2}{2}     \frac{\partial^2 }{\partial v^2}p(t-a,v)   -\frac{\partial }{\partial v}[(\mu -v)p(t-a,v)]  \right) \dd v.
 \end{split}
\end{equation*}
On the other hand, the second term in the equation can be written equivalently as
\begin{equation*}
\begin{split}
&\int_{-\infty}^1 \psi (a,v) \frac{\partial }{\partial a}  p(t-a, v) \dd v  =
\int_{-\infty}^1 \psi (a,v) \left(- \frac{\partial }{\partial t}  p(t-a, v) \right) \dd v, \\
\end{split}
\end{equation*}
which implies that
\begin{equation*}
\begin{split}
&\int_{-\infty}^1 \psi (a,v) \frac{\partial }{\partial a}  p(t-a, v) \dd v\\
& = \int_{-\infty}^1 \psi (a,v) \left(
       \frac{\partial }{\partial v}[(\mu -v)p(t-a,v)] - \frac{ \sigma ^2}{2} \frac{\partial^2 }{\partial v^2}p(t-a,v) -
\delta (v-v_{r})r(t-a) \right) \dd v.
\end{split}
\end{equation*}
Thus, by adding the two terms, we finally get something expected
\begin{equation*}
- \frac{\partial }{\partial a}\int_{-\infty}^1 \psi (a,v) p(t-a, v) \dd v  =  \psi (a,v_r)r(t-a),\\
\end{equation*}
expression that can also be written, due to (\ref{eql}), as
\begin{equation*}
\begin{split}
- \frac{\partial }{\partial a}\int_{-\infty}^1 \psi (a,v) p(t-a, v) \dd v  =  r(t-a) \int_{-\infty}^1 \varphi (a,v) \dd v
=\mathcal{P}(a)r(t-a).\\
\end{split}
\end{equation*}
Since, as reminded at the beginning of the proof, the right-hand side of the above relation expresses the solution
$n(t,a)$ for the case $t>a$, the proof of the first statement is  complete.

To prove (\ref{Inverse02}), we start by reminding that, in the case $a\geq t$, integration of the
solution to the AS system gives:
\begin{equation*}
n(t,a)=\frac{\mathcal{P}(a)}{\mathcal{P}(a-t)} n_0(a-t),\quad a\geq t.
\end{equation*}
With this respect, given an initial density of ages, the solution for $a\geq t$ is completely determined
by the initial density $n_0$ and the knowledge of $\varphi$, since, as reminded,
\begin{equation*}
\mathcal{P}(a)=\int_{-\infty}^1 \varphi(a,v)\dd a .
\end{equation*}
We wish though to relate the solution $n$ for this case to the initial density of potentials, and to do so, we will
see that, as anticipated, the sole relation (\ref{C17060}) is sufficient.
Note for the beginning that the functions $\psi$ and $\varphi$ are adjoint; let us consider next the integral
\be
\int_{-\infty}^1 \psi(t,v)\varphi(a-t, v)\dd v.
\ee
One can show that the above integral does not depend explicitly of $t$ by computing
\be\label{o}
\frac{\partial}{\partial t}\int_{-\infty}^1 \psi(t,v)\varphi(a-t, v)\dd v
=\int_{-\infty}^1\left(\frac{\partial}{\partial t}\psi(t,v)\varphi(a-t, v)
+\psi(t,v)\frac{\partial}{\partial t}\varphi(a-t, v)\right)\dd v.\\
\ee
Since, due to the fact that $\psi$ is solution to backward CK equation, one can write down
\bd
\int_{-\infty}^1\frac{\partial}{\partial t}\psi(t,v)\varphi(a-t, v)\dd v
=\int_{-\infty}^1\varphi(a-t, v)\left((\mu-v)\frac{\partial}{\partial v}\psi((t,v)
+\frac{\sigma^2}{2}\frac{\partial^2}{\partial v^2}\psi(t,v)\right)\dd v,
\ed
integrating by parts in the right-hand side and using the boundary conditions, it follows that
\bda
&&\int_{-\infty}^1\frac{\partial}{\partial t}\psi((t,v)\varphi(a-t, v)\dd v\\=
&&\int_{-\infty}^1\psi(t,v)\left(\frac{\partial}{\partial v}[-(\mu-v)\varphi(a-t, v)]
+\frac{\sigma^2}{2}\frac{\partial^2}{\partial v^2}\varphi(a-t, v)\right)\dd v\\
=&&-\int_{-\infty}^1\psi(t,v)\frac{\partial}{\partial t}\varphi(a-t, v)\dd v.
\eda
Replacing therefore the last expression in (\ref{o}), we get indeed that
\bd
\frac{\partial}{\partial t}\int_{-\infty}^1 \psi(t,v)\varphi(a-t, v)\dd v=0,
\ed
which implies that the above integral is only a function of $a$ on the interval $t\in [0,a]$.
We obtain therefore, due to initial conditions  $\varphi(0,v)=\delta(v-v_r)$ and $\psi(0,v)=1$, that
\bd
\left.\int_{-\infty}^1 \psi(t,v)\varphi(a-t, v)\dd v\right|_{t=0}=\int_{-\infty}^1 \varphi(a, v)\dd v
=\left.\int_{-\infty}^1 \psi(t,v)\varphi(a-t, v)\dd v\right|_{t=a}.
\ed
Going back to the expression of the solution for this case, one can write down then equivalently
\bd
n(t,a)=\frac{n_0(a-t)}{\mathcal{P}(a-t)}\int_{-\infty}^1 \psi(t,v)\varphi(a-t,v)\dd v ,\quad a\geq t,
\ed
and, integrating the last relation over $[t,\infty)$:
\bd
\int_{t}^\infty n(t,a)\dd a=
\int_{t}^\infty\frac{n_0(a-t)}{\mathcal{P}(a-t)}\int_{-\infty}^1 \psi(t,v)\varphi(a-t,v)\dd v \dd a.
\ed
Making the change of variable $a\lra a'=a-t$ and changing the order of integration, we finally get
\bd
\int_{t}^\infty n(t,a)\dd a =\int_{-\infty}^1 \psi(t,v)\int_0^\infty\frac{\varphi(a,v)}{\mathcal{P}(a)}n_0(a)\dd a\dd v
\ed
which leads us, due to (\ref{C17060}), to
\bd
\int_{t}^\infty n(t,a)\dd a=\int_{-\infty}^1 \psi(t,v)p_0(v)\dd v,
\ed
which ends the proof.
\end{proof}

\begin{remark}
In \cite{usdt} it was shown that, even if the relation between the initial states (\ref{C17060}) is not satisfied,
nevertheless, (\ref{C1706}) is satisfied asymptotically as $t\ra +\infty$. In a similar way,
Theorem \ref{Inverse} shows that even if $n(t,a)$ is well defined only for $0<a<t$, the relation (\ref{Inverse02})
assures that the influence of the undetermined part of the solution for $a>t$ goes to zero as $t\ra\infty$.
\end{remark}

 \begin{remark}
 Given the interpretations of all the functions involved in the
relation (\ref{Inverse01}), one can interpret the result probabilistically in the following way:
The probability density for a
neuron to have the age $a$ at time $t$ is given by the
flux generated by the probability density at potential $v$ and time $t-a$ that survived up to age $a$.\\
 The left-hand side in (\ref{Inverse02}) has the interpretation of the probability that the cell has at time $t$ an age
$a$ belonging to $ [t,+\infty)$.
With this respect, the relation simply says that, under assumption
(\ref{ir}),  the integral over $v$ of the probability density at time $0$  which
survived up to $t$ without firing
gives indeed the probability that $a > t$.

\end{remark}

\begin{remark}
The solution to the AS system defined in Theorem \ref{Inverse}
is indeed a probability density, i.e.

\be
\int_0^\infty n(t,a)\dd a= 1.
\ee
This is easily obtained by noting that
\bda
&&\int_0^\infty n(t,a)\dd a=\int_0^t n(t,a)\dd a+\int_t^\infty n(t,a)\dd a\\
&&=\int_0^t \left[-\frac{\partial}{\partial a} \int_{-\infty}^1 \psi(a,v)p(t-a,v) \dd v\right]\dd a+ \int_{-\infty}^1 \psi(t,v)p_0(v)\dd v\\
&&=1-\int_{-\infty}^1 \psi(t,v)p_0(v)\dd v+\int_{-\infty}^1 \psi(t,v)p_0(v)\dd v=1
\eda
\end{remark}

\section{Conclusions}

How to give a good analytical treatment to neural variability? Theories abound and provide very different ways to deal with the
stochastic aspect of nervous cells. Among them, probably the most popular approaches are the NLIF model with its
associated FP equation and the escape rate model
with its corresponding AS system. The FP equation is commonly used to represent the dynamics of the probability density function with respect to
the membrane potential and has the advantage that the parameters of the system have bio-physical interpretations and could be
measured/identified by experimentalists. Anyway, finding the membrane's potential density is rarely of use in practice.
In particular, the model does not give any insight view about the connection with previous spike times and the density of the
firing times. Models that give the evolution of a density with respect to the time elapsed since the last spike time (AS systems)
have been considered in the literature: \cite{gerstner,perthame03,perthame04}.
 These models though, in reverse, are not
linked to the bio-physical properties of the cell since they rely only on the assumption that a neuron fires
with the probability given by the escape rate.
The use of escape rate instead of consideration of diffusive noise has another obvious practical advantage:
while FP equation (\ref{model}) along with its boundary conditions and the source term given by (\ref{firingP}) is difficult to be handled from a mathematical point of view, the AS model (\ref{age}) - (\ref{id}) can be easily integrated
and leads to an analytical solution. This model is well known in the age-structured systems theory and a large
amount of papers have treated similar systems.
We mention that qualitative results for a similar model in neuroscience context have been obtained in \cite{perthame03,perthame04}.

It has been shown in \cite{plesser} that the NLIF model  can be mapped approximately
onto an escape-rate model.
In a recent study \cite{usdt}, we have shown an analytical connection between the two models.
We have proven there the existence of an exact analytical transform of the solution of the FP system into the solution to the AS
system. To our knowledge, such a result has not been proven before.
The present paper is intended as a sequel step toward the completion of the analytical link between these two known models.

The importance of our analytical connection consists in the fact that it gives a way to rely the density of ages to those of
 membrane potentials and underline in which way these densities depend on each-other; it is therefore, a first attempt to
 attribute age to potentials densities, information that is not carried out in the solution of the FP equation.
 Reversely, a more practical aspect of our transforms consists in the fact that it allows a  simpler analysis of the properties
 of the solutions by noticing in which way their properties transfer one to another via the integral transforms proposed here.
 Finally, from a biological point of view,
 our theoretical result shows that the two mathematical formalization of variability are similar.

We have to stress though that the results obtained here have been proven in the case of time independent stimulus.
The case we considered is known in the framework of renewal systems as a stationary process.
A possible extension of the present work
for the case of time dependant stimulus remains thus for us an open issue to be investigated. Such consideration would allow us
to discuss the case of interconnected neurons and thus get a better understanding of neural networks' dynamics. This is a current
work in progress.

\bibliographystyle{plain}
\bibliography{Inverse_article}
\end{document}